\documentclass[epj]{webofc}
\usepackage[varg]{txfonts}   
\woctitle{XLVI International Symposium on Multiparticle Dynamics}
\begin{document}
\title{Saturation and geometrical scaling }
%
%

\author{Michal Praszalowicz\inst{1}\fnsep\thanks{\email{michal@if.uj.edu.pl}}}

\institute{M. Smoluchowski Institute of Physics,
           Jagiellonian University,
           Lojasiewicza 11, 30-348 Krakow, Poland
          }

\abstract{%
We discuss emergence of geometrical scaling as a consequence of the non-linear evolution equations of QCD,
which generate  a new dynamical scale, known as the saturation momentum: $Q_{\rm s}$. In the kinematical region where
no other energy scales exist, particle spectra exhibit geometrical scaling (GS), {\em i.e.} they depend on the ratio 
$p_{\rm T}/Q_{\rm s}$, and the energy dependence enters solely through the energy dependence of the
saturation momentum.
We confront the hypothesis of GS in different systems with experimental data.
}
\maketitle
\section{Introduction}
\label{intro}
In this report we present a concise analysis of GS, slightly extended with respect to the  presentation
given at the XLVI International Symposium on Multiparticle Dynamics. One can find more details
in the original publications \cite{Praszalowicz:2012zh}\nocite{McLerran:2010ex,Praszalowicz:2011rm}--
\cite{Praszalowicz:2015dta} and in the recent conference proceedings \cite{Praszalowicz:2016jcg},
that cover the same topics.

In QCD we have basically two sets  of evolution equations that describe the change of parton densities with
decreasing resolution scale $1/Q^2$ -- DGLAP equations, or with growing energy (or equivalently with
decreasing Bjorken $x$) -- BFKL equation. In both cases the number of partons, or more precisely 
 the number of gluons, is growing rapidly with the evolution variable. In the BFKL case however (since the
average transverse size of gluons is fixed), we enter a regime where the partonic system is not dilute and the
linear evolution breaks down. A modified  BFKL equation that includes the non-linear terms is known as
the Balitsky-Kovchegov (BK) equation \cite{BK}. One of the consequences of the nonlinearities is the emergence of
the so called saturation scale of the form  \cite{Mueller:2002zm,Munier:2003vc}:
\begin{equation}
Q_{\text{s}}^{2}(x)=Q_{0}^{2}( {x}/{x_{0}})
^{-\lambda} .
\label{Qsat}
\end{equation}

Munier and Peschanski \cite{Munier:2003vc} draw an analogy between the BK equation and the time evolution of the wave front
$u(t,z)$ in one dimensional space variable $z$:
\begin{equation}
\frac{\partial}{\partial t} u(t,z) = {\cal O}(\partial /\partial z)
\label{traveleq}
\end{equation}  
where ${\cal O}(\partial /\partial z)$ is a non-linear differential operator corresponding in QCD to the BK kernel.
For a wide class of operators ${\cal O}$ and initial conditions for $u$, wave front $u$ converges 
asymptotically to the {\em traveling wave}:
\begin{equation}
u(x,z) \rightarrow u(z-v_{\rm c} t),
\label{travelsol}
\end{equation}
{\em i.e.} to the fixed front that moves rigidly with the critical velocity $v_{\rm c}$. In the QCD
context time corresponds  to the logarithm of Bjorken $x$: $t =\ln (x_0/x)$ where $x_0$ is
a constant  corresponding to $t=0$, and $z=\ln (p_{\rm T}^2/Q_0^2)$.
Translating the argument of the travelling wave to the QCD variables gives:
\begin{equation}
z-v_{\rm c} t=\ln \left( \frac{p_{\rm T}^2}{Q_0^2}\right)-v_{\rm c} \ln \left(\frac{x_0}{x} \right)
=\ln \left( \frac{p_{\rm T}^2}{Q_0^2 (x/x_0)^{-v_{\rm c}}}\right).
\label{travelQCD}
\end{equation} 
Therefore the travelling wave corresponds to the scaling solution
with the saturation momentum given by Eq.~(\ref{Qsat}).
In the following we shall check whether GS is present in different pieces of high energy data.

\section{Deep inelastic scattering (DIS)}
\label{sec:dis}

\begin{figure}[h]
\centering
\includegraphics[width=7cm,angle=0]{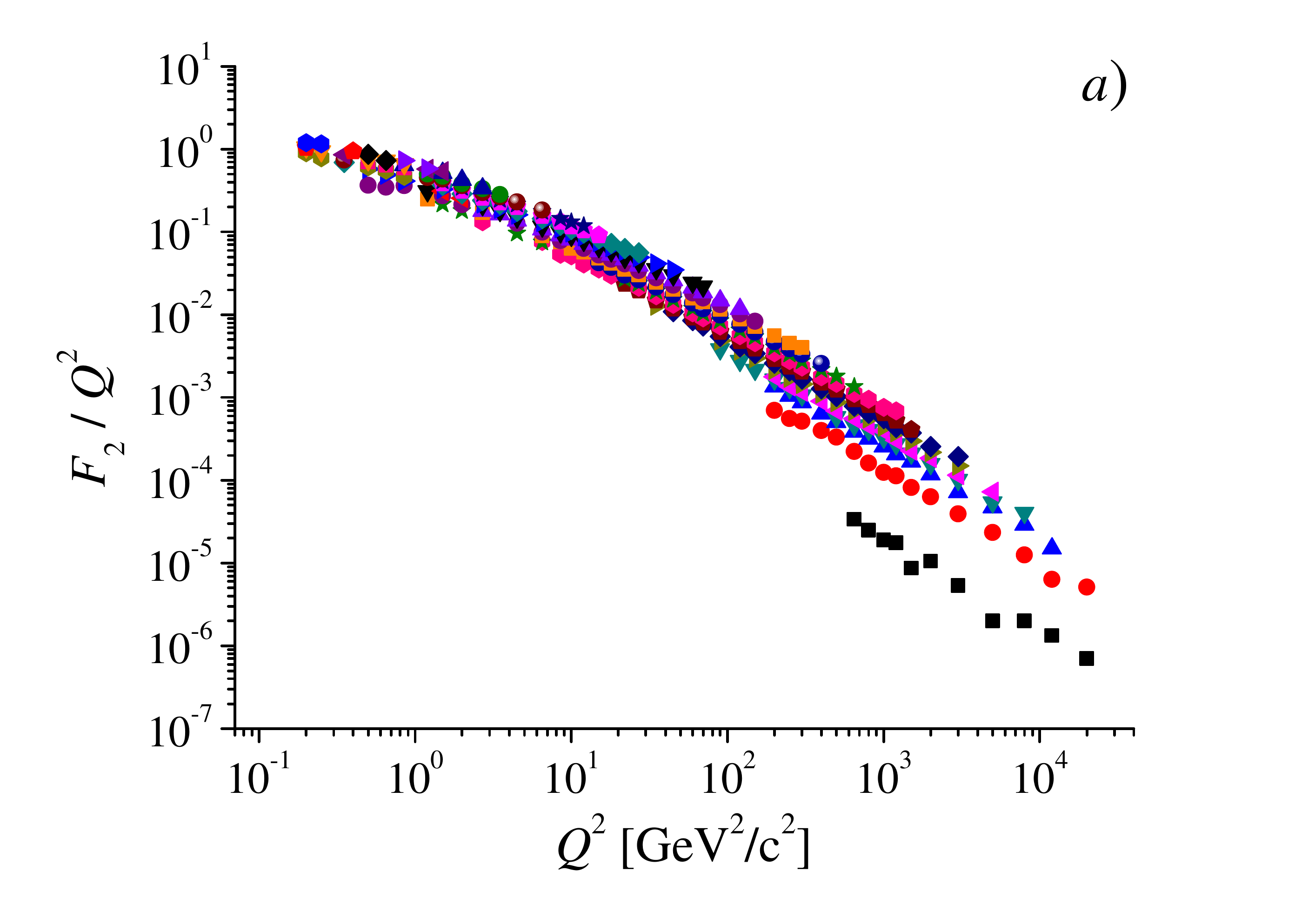}
\includegraphics[width=7cm,angle=0]{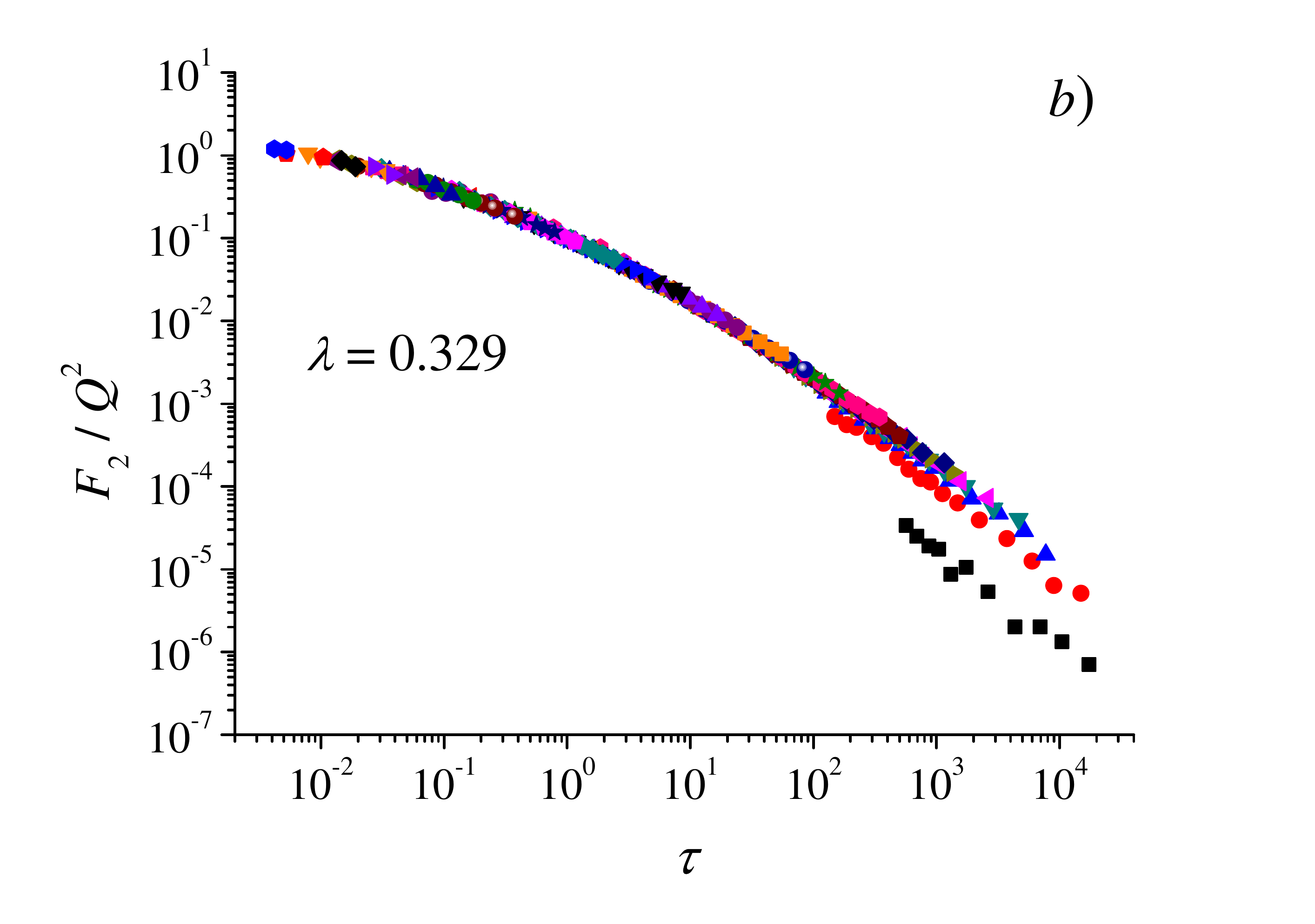} \caption{Combined DIS data
\cite{HERAdata} for $F_{2}/Q^{2}$. Different points forming a wide band as a
function of $Q^{2}$ in the left panel correspond to different Bjorken $x$'s.
They fall on a universal curve when plotted in terms of $\tau$ (right panel).
(Figure from the first paper of Ref.~\cite{Praszalowicz:2012zh}).}%
\label{GSDIS}%
\end{figure}

Geometrical scaling was first introduced in the context of DIS for $F_{2}%
(x)/Q^{2}$~\cite{Stasto:2000er}. In Fig.~\ref{GSDIS} we plot $F_{2}(x)/Q^{2}$
as a function of $Q^{2}$ (left panel) and in terms of the scaling variable $\tau=Q^2/Q_{\rm s}^2 (x)$ for
$\lambda=0.329$ (right panel) for the combined HERA data \cite{HERAdata}.
Points of different colors correspond to different Bjorken $x$'s. We see from
Fig.~\ref{GSDIS} that DIS data scale very well with
some exception in the right part of Fig.~\ref{GSDIS}.b. These points, however,
correspond to large Bjorken $x$'s where GS is supposed to break.

Since the reduced cross-section in ep scattering (which is essentially proportional to $F_{2}/Q^{2}$)
is given as a convolution of the virtual photon wave function and an {\rm unintegrated gluon distribution} of
the proton $\varphi_{p}\left(  {\vec{k}_{\text{T}}^{\, 2}},x \right)  $, the fact the DIS data scale implies that
$\varphi_{p}\left(  {{k}_{\text{T}}^{\, 2}},x \right)  =\varphi_{p}\left(  {{k}_{\text{T}}^{\, 2}}/Q^2_{\rm s}(x) \right)$.

\section{Inelastic $p_{\mathrm{T}}$ spectra at the LHC}
\label{ppLHC}

The cross-section for not too hard gluon production in pp
collisions can be described in the $k_{\text{T}}-$factorization approach by
the formula \cite{Gribov:1981kg}:%
\begin{equation}
\frac{d\sigma}{dyd^{2}p_{\text{T}}}=\frac{3\pi\alpha_{\text{s}}}{2 p_{\text{T}}^{2}}
{\displaystyle\int}
{d^{2}\vec{k}_{\text{T}}}\,
\varphi_{p}\left(  {\vec{k}_{\text{T}}^{\, 2}},x_1 \right)  
\varphi_{p}\left(  {(\vec{k}-\vec{p}\,)_{\text{T}}^{2} , x_2}\right) \label{sigma_1}%
\end{equation}
where $\varphi_{p}$ denotes the unintegrated gluon distribution. 
Here $x_{1,2}=e^{\pm y}p_{\rm T}/\sqrt{s}$.
In the  following
we shall assume that produced gluons
are in the mid rapidity region ($y\simeq0$), hence both Bjorken $x$'s of
colliding gluons are equal $x_{1}\simeq x_{2}$ (denoted in the following as
$x$). 

Unintegrated gluon distributions have dimension of transverse area. We shall assume that -- following
our discussion of DIS -- that they depend only on the ratio of the transverse momentum over the saturation scale
\begin{equation}
\varphi_{p}\left(  {k}_{\text{T}}^{\, 2},x \right) = S_\bot \phi \left({k}_{\text{T}}^{\, 2}/Q_{\rm s}^2(x) \right)
\end{equation}
where $\phi$ is a {\em dimensionless} function. 
In the case of DIS $S_{\bot}=\sigma_{0}$ is the dipole-proton cross-section
for large dipoles \cite{GolecBiernat:1998js}. In the case of heavy ion collisions  $S_{\bot}$ is the transverse size of
an overlap of two large nuclei for a given centrality class. In both cases one
can assume that $S_{\bot}$ is energy independent (or weakly dependent). In that case $d^{2}\vec{k}_{\text{T}}$
integration in (\ref{sigma_1}) leads to%
\begin{equation}
\frac{d\sigma}{dyd^{2}p_{\text{T}}}=S_{\bot}^{2}\mathcal{F}(\tau
)\label{sigma_2}%
\end{equation}
where $ \tau=p_{\rm{T}}^{2}/Q_{\rm s}^{2}(x)$ is a scaling variable
and $\mathcal{F}(\tau)$ is a function related to the integral of $\phi_{p}%
$'s. We follow here the parton-hadron duality hypothesis
 \cite{PHD}, assuming that the charged particle spectra are on the
average identical to the gluon spectra.

\begin{figure}[h]
\centering
\includegraphics[width=6cm,angle=0]{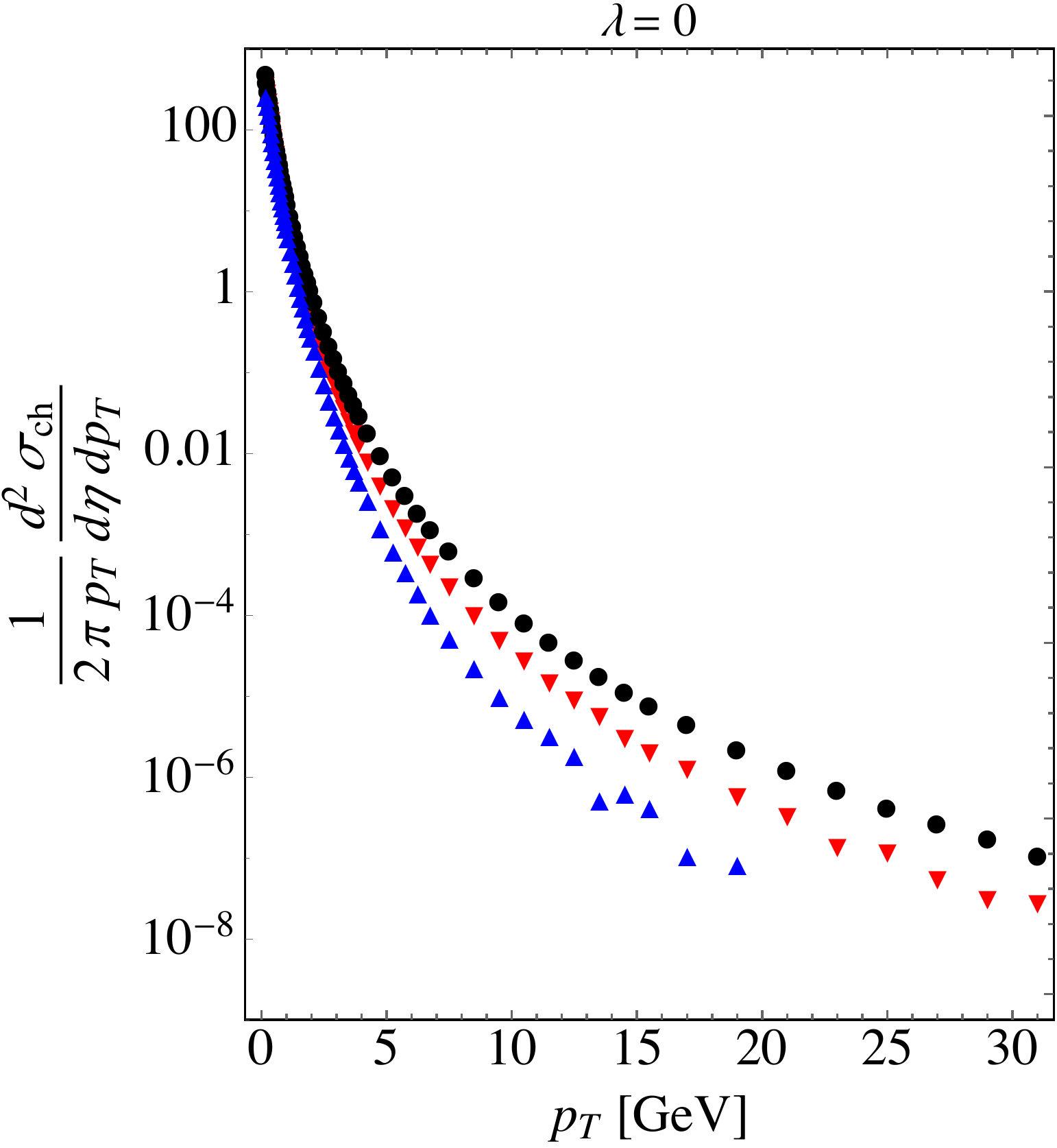}
\includegraphics[width=6cm,angle=0]{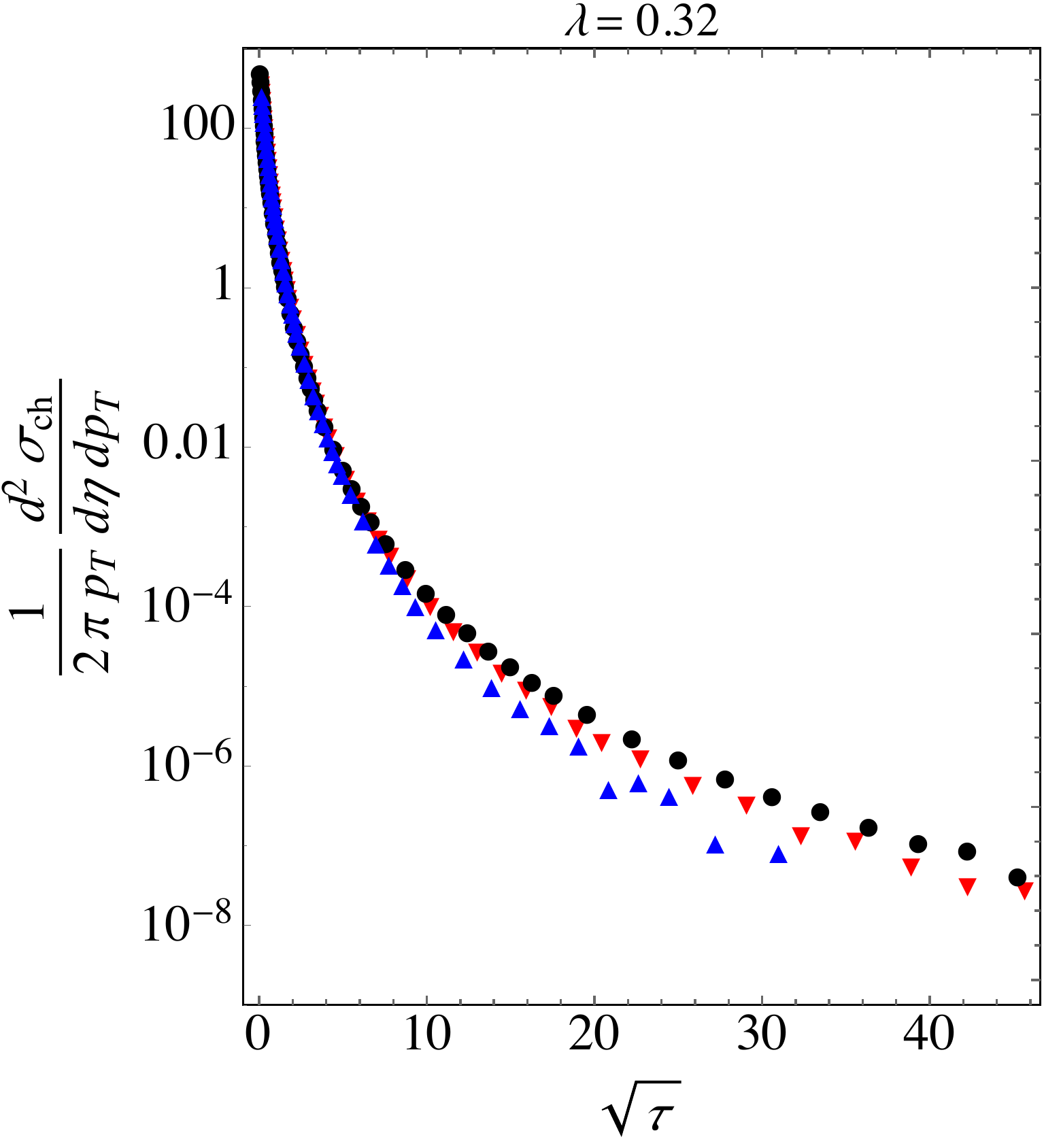}\caption{Data for
pp scattering from ALICE \cite{Abelev:2013ala} plotted in terms of
$p_{\mathrm{T}}$ and $\sqrt{\tau}$. Full (black) circles correspond to
$W=7$~TeV, down (red) triangles to 2.76~TeV and up (blue) triangles to
0.9~TeV.
(Figure from Ref.~\cite{Praszalowicz:2015dta}.)}%
\label{GSALICE}%
\end{figure}

In
Fig.~\ref{GSALICE} we plot ALICE pp data \cite{Abelev:2013ala} in terms of
$p_{\mathrm{T}}$ (left panel) and in terms of the scaling variable $\tau$ (right
panel) for $\lambda=0.32$. We see that three different curves from the left
panel in Fig.~\ref{GSALICE} overlap over some region if plotted in terms of
 $\sqrt{\tau}$. The exponent for which this happens over the
largest interval of $\tau$ is $\lambda=0.32$ \cite{Praszalowicz:2015dta}, 
which is the value compatible
with our model independent analysis of the DIS data \cite{Praszalowicz:2012zh}.

\begin{figure}[t]
\centering
\includegraphics[width=6cm,angle=0]{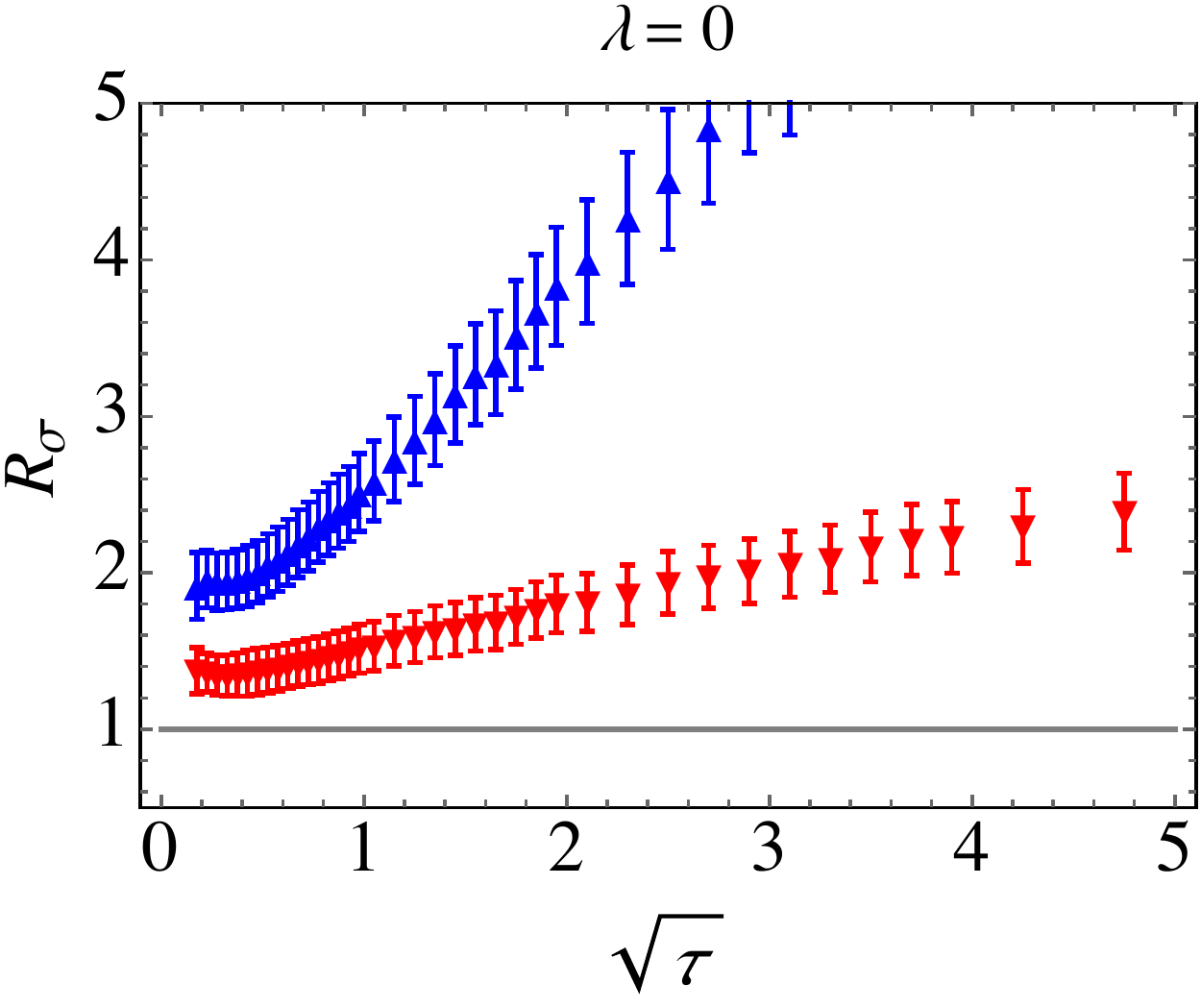}
\includegraphics[width=6cm,angle=0]{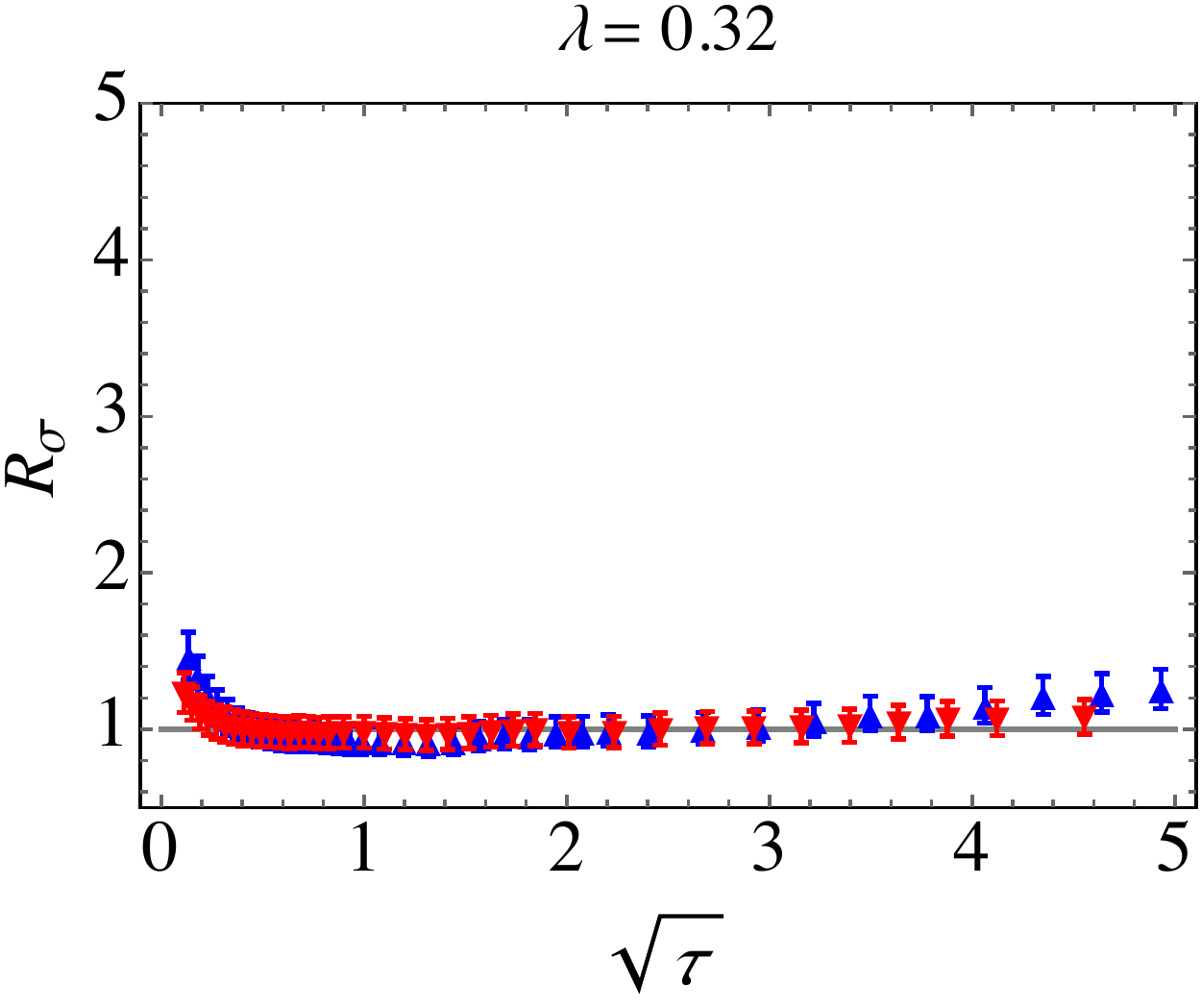}\caption{Ratios
of the cross-sections at 7/2.76 TeV -- down (red) triangles and 7/0.9 TeV -- up (blue) triangles,
for  $\lambda=0$ (left) and 0.32 (right). (Figure from Ref.~\cite{Praszalowicz:2015dta}.)}%
\label{Rsigma}%
\end{figure}

In order to illustrate the method of adjusting $\lambda$, we plot in Fig.
\ref{Rsigma} ratios of the cross-sections at 7 TeV to 2.76 and 0.9 TeV.
Proximity of both ratios to unity for $\lambda=0.32$ 
is the sign of GS for
$p_{\text{T}}$ up to $4.25$ GeV$/c$ \cite{Praszalowicz:2015dta}.

It has been argued previously \cite{McLerran:2010ex} that GS should hold for multiplicities, rather
than for the cross-sections. This would be true if the relation between the
two was energy independent. This may be the case in HI or p$A$ collisions where we
trigger on some $S_{\bot}$ by selecting the centrality classes with given
number of participants, but it is not true in the case of the inelastic pp
scattering:
\begin{equation}
\frac{dN}{dyd^{2}p_{\text{T}}}=\frac{1}{\sigma^{\mathrm{MB}}(W)}\frac{d\sigma
}{dyd^{2}p_{\text{T}}}=\frac{S_{\bot}^{2}}{\sigma^{\mathrm{MB}}(W)}%
\mathcal{F}(\tau)
\end{equation}
where the minimum bias cross-section $\sigma^{\mathrm{MB}}(W)\neq S_{\bot}$ is
energy-dependent. Nevertheless the multiplicity does scale, however for different value
of the exponent  ($\lambda=0.22 - 0.24$), and over a smaller range of $p_{\rm T}$.
This is discussed in more detail in Ref.~\cite{Praszalowicz:2015dta}.

\section{Electron-positron scattering}
\label{e+e-}

One of the obvious questions concerning pp scattering is whether the observed scaling is indeed
a consequence of gluon saturation or whether it appears due to the properties of final state radiation
that takes place before hadronization (note that by referring to the parton-hadron duality we have ignored
possible effects of the QCD fragmentation functions). The best way to address this issue is to perform
analogical analysis of the hadronic $p_{\rm T}$ spectra in e$^+$e$^-$ collisions where no GS is expected.
To this end we shall use spectra published by TASSO collaboration \cite{Braunschweig:1990yd}
at $W=\sqrt{s}=14$, 22, 35 and 44 GeV.\footnote{The author is grateful to J. Chwastowski for pointing 
 this reference.}
One might argue that these energies are too small to reach a firm conclusion on GS in e$^+$e$^-$. However,
in Ref.~\cite{Praszalowicz:2013uu} we have analyzed pp scattering at even lower energies (6 -- 17 GeV) 
finding positive evidence of GS
in the mid rapidity region.

\begin{figure}[t]
\centering
\includegraphics[width=6cm,angle=0]{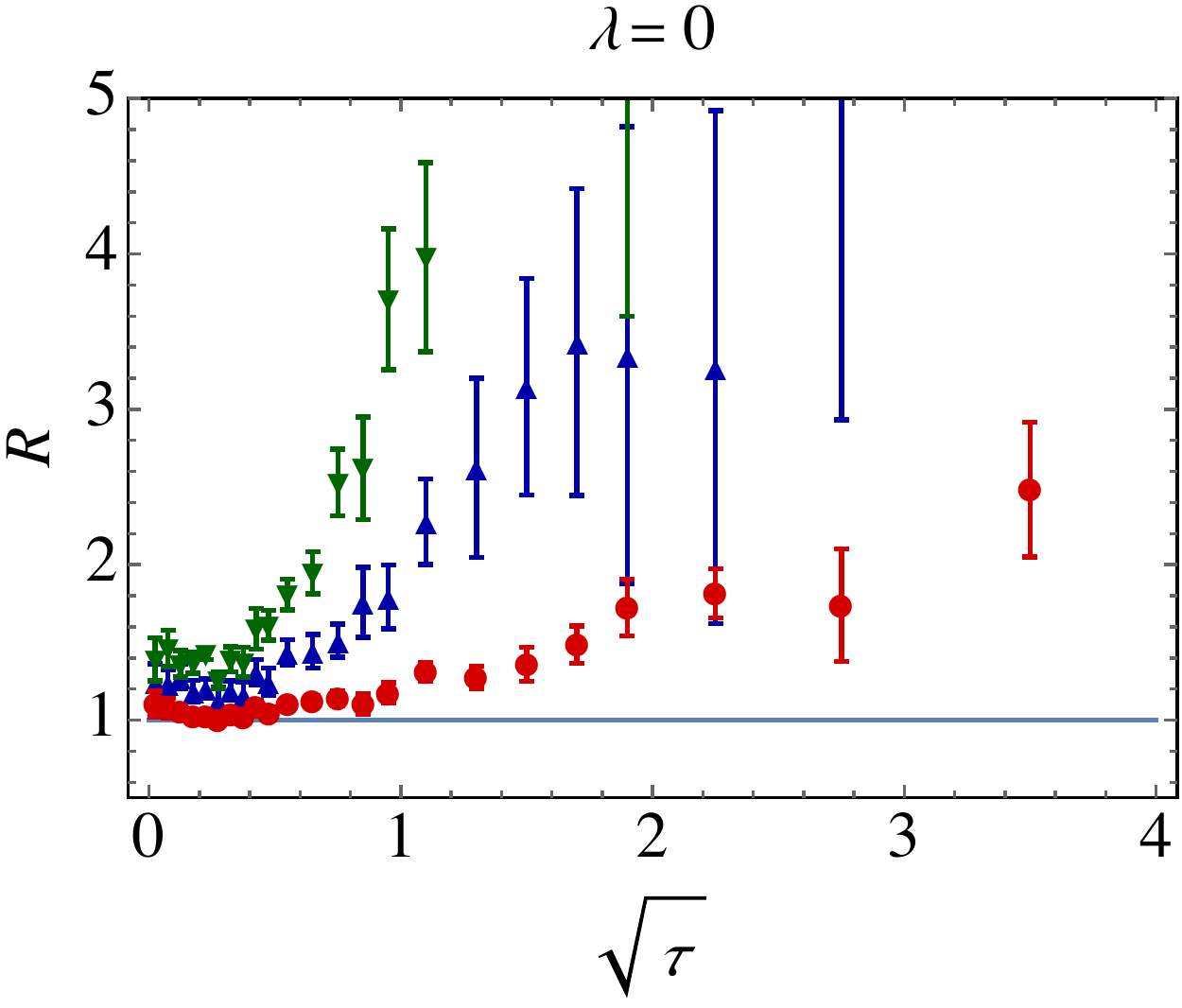}
\includegraphics[width=6cm,angle=0]{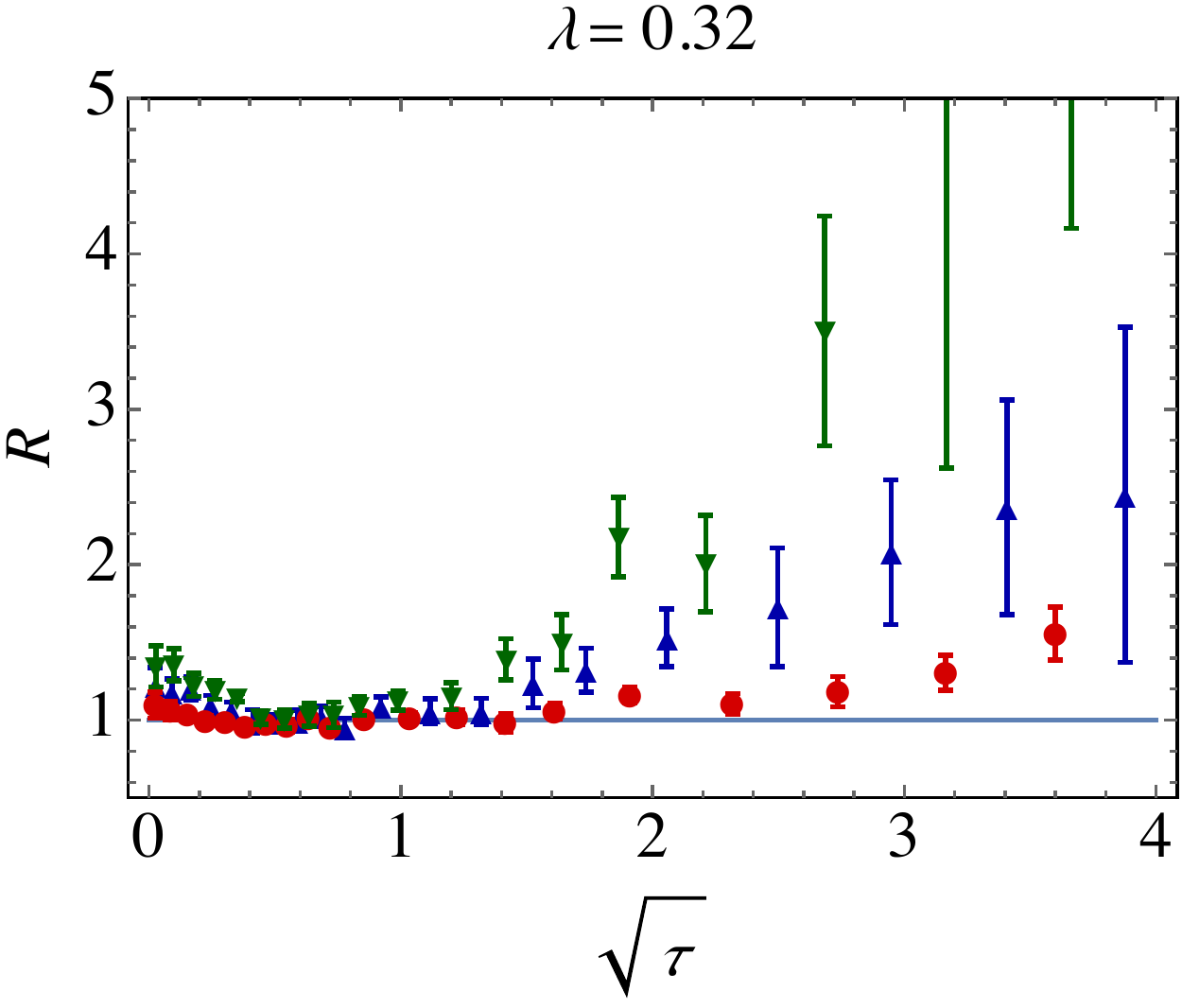}\caption{Ratios $R(W)$
of multiplicity spectra for $W=14$ (down-green triangles), 22 (up-blue triangles) and 35 GeV (red-full circles) as 
functions of $p_{\rm T}$ (left panel) and $\sqrt{\tau}$ for $\lambda=0.32$ (right panel).
}%
\label{RTASSO}%
\end{figure}

In Fig.~\ref{RTASSO} ratios  of multiplicity spectra 
\begin{equation}
R(W)=\left. \frac{1}{\sigma_{tot}}\frac{d\sigma}{2\pi p_{\rm T} d p_{\rm T}}\right|_{44~{\rm GeV}}
{\mbox{\LARGE /}}
\left. \frac{1}{\sigma_{tot}}\frac{d\sigma}{2\pi p_{\rm T} d p_{\rm T}}\right|_{W}
\end{equation}
for the remaining three values of $W$ are plotted in terms of the scaling variable $\sqrt{\tau}$ for $\lambda=0$
(note that in this case $\sqrt{\tau}=p_{\rm T}$) and for $\lambda=0.32$. In the left panel w see similar behaviour
of $R$ as in the pp case (see Fig.~\ref{Rsigma}.a), however in the right panel where we plot $R$ for $\lambda=0.32$
no GS is seen (compare with  Fig.~\ref{Rsigma}.b).

\section{Summary}
\label{sum}

In this short note we have argued that geometrical scaling is clearly seen in the DIS and pp data. In the latter case
an observable that scales is the differential cross-section rather than the multiplicity distribution. Phenomenological
analysis analogical to the one described here has been used to look for the effects of GS in p$A$ and heavy ion collisions
and in $\langle p_{\rm T} \rangle$ correlation with multiplicity \cite{McLerran:2010ex}--\cite{Praszalowicz:2016jcg}.
Most recently effects of the fluctuations of the saturation momentum in the p$A$ collisions have been studied
in Ref.~\cite{McLerran:2015lta}.

Scaling violations that emerge in the kinematical limit when one of the Bjorken $x$'s in (\ref{sigma_1})  is close to unity,
have been observed
in Ref.~\cite{Praszalowicz:2013uu}. Here, in Sect.~\ref{e+e-}, we have presented a new evidence that GS is not present 
in the spectra of particles produced in e$^+$e$^-$ collisions. Although this has been theoretically expected, conclusions
drawn from Fig.~\ref{RTASSO} support our main hypothesis that geometrical scaling seen in the experimental data
is due to the saturation effects in the initial state of the colliding hadrons. 

\section*{Ackowledgements}
The author wants to thank the organizers  for an invitation and for putting up together this successful
meeting. This work was supported by the Polish NCN grant 2014/13/B/ST2/02486.

%
%

\end{document}